\documentstyle[11pt]{article}
\title{~\\~\\~\\~\\
 TECHNICOLOR ENHANCEMENT OF $t \bar{t}$ PRODUCTION AT TeV-COLLIDERS}
\author{ Thomas Appelquist and George Triantaphyllou\\
Department of Physics, Yale University, New Haven, Ct. 06520}
\begin{document}
\setlength{\baselineskip}{24pt}
\maketitle

\begin{abstract}
It is shown that a technicolor theory  containing a color-octet
technipion, usually denoted by $P^{0'}_{8}$, will give rise to an
enhancement of $t \bar t$
production at the Tevatron, LHC and SSC, via the process
$gg \rightarrow P^{0'}_{8} \rightarrow
t \bar t$. The relevant cross-sections
are computed taking into account the large lower bound on the top mass
coming from the "top search" experiments at LEP and CDF.
 At the LHC and SSC, the signal
is  found to be comparable to the QCD background, making the process
quite accesible.
\end{abstract}
\setcounter{page}{0}
\pagebreak

Technicolor theories typically contain pseudo-Goldstone bosons
 (technipions), which arise
from the breakdown of global chiral symmetries. Here,  the
production and subsequent decay of a color-octet technipion at the
Tevatron, LHC and SSC  is studied. The sub-process considered is
$\;\; {\rm\ gluon-gluon} \rightarrow P^{0'}_8 \rightarrow t \bar{t} $,
where $P^{0'}_8$ is the
technipion. The subscript "8" denotes that
it is a QCD octet, the superscript "0" that it is neutral,
and the prime that it is a singlet under the weak $SU(2)_{L}$.
This particle is expected to be light compared to most technihadrons, and
therefore more accessible to the TeV-colliders.
However, it is assumed
throughout this paper that its mass is above the
$t\bar{t}$ production threshold.

The effective ETC couplings of the technipions to the quarks are
proportional to $ m_{q}/F_{\pi}$. Therefore,
  $q \bar{q}$ fusion as a production mechanism is not considered,
  nor is the  decay of the
technipions into quarks lighter than the top or the bottom.   The
decay of the technipions to two bottom quarks or to two gluons
is not considered either, even though it gives non-negligible cross-sections,
 since the background to these
processes makes them difficult to observe.
Moreover, we consider only the color-octet $P^{0'}_{8}$
as the intermediate technipion,
 because color counting factors make its
cross-section approximately eight times larger than the color-singlet one.
Also, because the color-singlet technipion mass does not
receive a QCD contribution, it could lie below the
$t\bar{t}$ production threshold. The results of this paper
  could, however, also be applied to a color-singlet technipion,
 if corrected by the relevant counting factors.

Color-octet technipions exist only in
models with at least one colored doublet of technifermions.
 A one family model will be employed here.
It has  a global $ SU(8)_{L} \times SU(8)_{R}$ chiral symmetry that breaks
down to an $SU(8)_{L+R}$ and thus produces 63 Goldstone bosons, 3 of
which are "eaten" by the $W$'s and the $Z^{0}$, and 60 of which acquire masses
depending on their quantum numbers.
Even though present experimental
constraints on the S parameter do not favor a
large technicolor sector \cite{ST},
a theory with a complete techni-family is viable if the technicolor group
is not too large \cite{AT}.
In the following, the technicolor gauge group
$SU(N_{TC})$ with $N_{TC} = 2$ will be used.

The process in question attracted the interest of
several authors some  years ago [4-7]. However, the top
mass used in those computations never exceeded $70$ GeV \cite{EHLQ}. Given the
dependance of the technicolor signal and the QCD background
on the quark and technipion masses,
 the current lower bound
on $m_{t}$ ( $m_{t} > 91$ GeV \cite{CDF}) and the emergence of new classes
of technicolor theories (the so-called walking theories \cite{walk})
that lead to larger technipion masses make it important to recompute
 the relevant cross-sections.

 The mass of the color-octet technipion is estimated first. It
receives contributions  mainly from QCD and
  ETC interactions. The QCD contribution, according to a previous estimate
   \cite{technimass}, gives
   \begin{equation}
    M^{2}_{P} \approx \frac{9 \ln 2}{2 \pi}
\alpha_{s}(M^{2}_{P})\;M^{2}_{V} \;\;,
   \label{eq:mass}
   \end{equation}

   \noindent where $M_{V}$ is the mass of the lightest technivector resonance.
To find $M_{V}$, the following scaling relation is used:
\begin{equation}
M_{V} = (3/N_{TC})^{1/2}(F_{\pi}/f_{\pi})\;m_{\rho}\;\;,
\end{equation}
where $f_{\pi}$ is the
pion decay constant, and
 $F_{\pi} = \frac{246 \;{\rm GeV}}{\sqrt{N_{D}}}$, with $N_{D}$
the number of technidoublets.   For the one family model, $N_{D} = 4$, so
  $F_{\pi} = 123$ GeV.
 Inserting the expression for $M_{V}$
into Eq.\ref{eq:mass} gives $M_{P} \approx 400$ GeV .

There is also  an ETC contribution to the $P^{0'}_{8}$ mass.
 A previous estimate gives roughly \cite{AppleRo}
   \begin{equation}
   M^{2}_{P} \approx  g^{2}_{ETC} \frac{<\psi\bar{\psi}>^{2}}
  {M^{2}_{ETC}F_{\pi}^{2}}\;\;,
   \end{equation}
\noindent where $M_{ETC}$ can range from 10 to 1000 TeV, and
$g^{2}_{ETC}/4\pi^{2}$ is expected to be {\em O}(1). In conventional
technicolor theories,
the chiral condensate $ <\psi\bar{\psi}> \approx
\Lambda^{3}_{TC}/4\pi^{2}\;$, where
$\Lambda_{TC} \approx \frac{M_{V}}{m_{\rho}} \Lambda_{QCD}$
denotes the technicolor confinement scale.
 In these theories, therefore,
the ETC contribution to the
technipion mass can be at most about 100 GeV.
Walking technicolor models, however, can give larger masses to the technipions
 via high-momentum enhancement \cite{walk}. The upper limit of the
technicolor condensate $ <\psi\bar{\psi}>$ in these models is of order
$M_{ETC}\Lambda^{2}_{TC}\;$, so the ETC contribution
could even approach 1 TeV.  The $SU(N_{TC} = 2)$ technicolor model with
$N_{D} = 4$ gives a somewhat enhanced condensate, but well
below the upper limit.
Masses in the range $350-550$ GeV will be
considered here.

Next, in order to determine the cross-section
for the sub-process $gg \rightarrow P^{0'}_{8} \rightarrow
t \bar t$,
the partial decay widths of the $P_{8}^{0'}$
 to two gluons and to a quark pair are needed. They are \cite{ggtt}
\begin{eqnarray}
  \Gamma(P^{0'}_{8} \rightarrow gg) & = & \frac{5N^{2}_{\,TC}}{384 \pi^{3}}\;
  \alpha_{s}^{2}(M^{2}_{P})
  \;\frac{M^{3}_{P}}{F_{\pi}^{2}} \;\;\;\; {\rm and}
\nonumber \\
& & \nonumber \\
  \Gamma(P^{0'}_{8} \rightarrow t \bar{t}) & \approx &
\frac{m^{2}_{t}M_{P}}{4 \pi
   F^{2}_{\pi}}\;\left(1-4 \frac{m^{2}_{t}}{M^{2}_{P}}\right)^{1/2}\;\;,
  \label{eq:widths}
  \end{eqnarray}
  \noindent  where $M_{P}$ is
the technipion  mass, and $\alpha_{s}^{2}(M^{2}_{P})$
is the QCD coupling evaluated
at $M^{2}_{P}$. The width for $t\bar{t}$ decay depends on
the specific assumption made for the ETC-induced
 coupling of the technipion to
 $t\bar{t}$. Here, a CP-conserving coupling
of strength $2m_{t}/F_{\pi}$ is used.
  A typical choice for the top mass, $m_{t} = 120$ GeV,
and the technipion mass, $M_{P} = 350$ GeV, gives
$0.05$ GeV and roughly $20$ GeV for
the decay widths of the techipion to $gg$ and $t\bar{t}$ respectively.
The latter could be smaller if the technipion mass
is closer to the $t\bar{t}$ production threshold.

To estimate the total decay width of the technipion, denoted here by
$\Gamma_{tot}$, we note that it decays predominantly into a
$t\bar{t}$ pair. This is  the case as long as
$m_{t}~^{>}_{\sim}\;2 \times 10^{-2} M_{P}$.
Thus, $\Gamma_{tot} \approx
\Gamma(P^{0'}_{8} \rightarrow t\bar{t})\;\;$. This estimate is  valid as long
as the technipion mass does not become very close the
$t\bar{t}$ production threshold, i.e.
as long as $1 - 4 m^{2}_{t}/M^{2}_{P} ~^{>}_{\sim} \;10^{-6}$.

Using the relativistic Breit-Wigner formula,
the sub-process cross-section, for $N_{\,TC} = 2$, then reads
   \begin{eqnarray}
   \hat{\sigma}_{\,TC}(gg\rightarrow P^{0'}_{8} \rightarrow t \bar{t}) & = &
    \frac{\pi}{2}
    \frac{\Gamma(P^{0'}_{8} \rightarrow gg)
   \;\Gamma(P^{0'}_{8} \rightarrow t \bar{t})}
   {(\hat{s} - M^{2}_{P})^{2}
   + M^{2}_{P} \Gamma^{2}_{tot}} \nonumber \\
   & & \nonumber \\
      & = &
   \frac{10}{3(8 \pi)^{3}}
   \frac{\alpha^{2}_{s}(M^{2}_{P})\;M^{4}_{P}\;m^{2}_{t}\;
   (1-4m^{2}_{t}/M^{2}_{P})^{1/2}}
   {F^{4}_{\pi}\;
    (\;(\hat{s} - M^{2}_{P})^{2} + M^{2}_{P} \Gamma^{2}_{tot}\;)}\;.
   \label{eq:crossect}
   \end{eqnarray}
\noindent The QCD final-state corrections are neglected in this
expression.
 The decay widths of Eq.\ref{eq:widths}
should actually be corrected before being inserted
in the cross-section for $gg \rightarrow P^{0'}_{8} \rightarrow t\bar{t}\;$
 since  the relevant quantity entering these widths
is not the mass of the technipion, but the
invariant mass $\sqrt{\hat{s}}$ of the process.
 These corrections are not taken into account, partly because
the use of strict invariant mass cuts makes their effects negligible, and
partly because the uncertainty introduced by the technipion and
top-quark masses, which are still experimentally unknown,
  makes such an analysis too detailed.
It should be noted in particular that Eq.\ref{eq:crossect} cannot be
applied for
 technipion masses, or $\sqrt{\hat{s}}$ values too close to
 the $t\bar{t}$ production threshold.

The result of Eq.\ref{eq:crossect} can be carried one step further,
in order to compute the
differential cross-section with respect to the scattering angle.
Since the technipion has zero spin,
 $\frac{ d \hat{\sigma}_{\,TC} } {d{ \rm cos} \theta} = \hat{\sigma}_{\,TC}/2$,
 where $\theta$ is the scattering angle measured in the
gluon-gluon center-of-mass frame.

The  ingredients are now in hand  to compute the
 integrated cross-section for
the process $pp \rightarrow P^{0'}_{8} \rightarrow
t \bar {t}\;$.
 Denoting by  $s$  the total C.M. energy of the $pp$
 beams, we have
 $\hat{s} = s x_{a} x_{b} = s \tau$, where $\tau \equiv x_{a} x_{b} \;$, and
 $x_{a}$ and $x_{b}$  are the momentum
fractions of the two interacting partons a and b, considered to be
massless. In our case, they are
gluons.
 The parton distribution functions for gluons, denoted by
$f_{g}(x_{i},\hat{s})$, $i = a,b$, can now be used,  to write
   \[ \sigma_{\,TC}(pp \rightarrow P^{0'}_{8} \rightarrow t \bar{t}) = \]
\begin{equation}
     = \int^{\tau_{+}}_{\tau_{-}}\!\! d\tau\!\!
\int^{Y}_{-Y}\!\! dy_{1}\int^{y_{+}}_{y_{-}}
\!\! dy_{2}\frac{f_{g}(x_{a},\hat{s})f_{g}(x_{b},\hat{s})}
         {2 \;\sqrt{1-4m^{2}_{t}/\hat{s}}\;{\rm cosh}^{2}
(\frac{y_{1}-y_{2}}{2})}
  \;\frac{ d \hat{\sigma}_{\,TC} } {d{ \rm cos} \theta}
(gg\rightarrow P^{0'}_{8} \rightarrow t \bar{t})\;,
   \label{eq:fincross}
   \end{equation}
   \noindent where  the $q \bar{q}$ contribution to the
 process has been neglected.
Here $y_{1,2}$ are the rapidities of the two top quarks
measured in the laboratory frame and $\tau$ is defined as above.
According to the above definitions,
$x_{~^{a}_{b}} = \sqrt{\tau}\,{\rm exp}( \pm \frac{y_{1}+y_{2}}{2})\;$.
 The quantities $y_{\pm}$ are defined as
$y_{\pm} = \;~^{min}_{max}(\pm Y,\mp \ln \tau - y_{1})$.
The  integration limits $\tau_{\pm},\,Y$, and $y_{\pm}$  correspond
to experimental cuts, and
 will be discussed shortly.
The integrated cross-section $\sigma_{\,TC}$ is then computed for various
top and technipion masses.
For the gluon distribution functions $f_{g}$,
the HMRS(B) functions are used, as described in Ref.\cite{parton}.
 The integrals in Eq.\ref{eq:fincross} are done numerically by a
Monte-Carlo algorithm \cite{BargKleis}.
The program is stopped as soon as $10^{4}$ Monte-Carlo-generated
events have passed all
the cuts. This gives  sufficiently
small statistical errors.

The form of Eq.\ref{eq:crossect} implies that, while the cross-section can
 increase quadratically with $m_{t}$, this
only happens for $m_{t}$ small enough so that the total width is dominated by
decays other than the decay $P^{0'}_{8} \rightarrow t\bar{t}$.
  For the allowed  $m_{t}$ range, however,
 $\Gamma(P^{0'}_{8} \rightarrow t\bar{t})$
dominates, making the integrated
cross-section (Eq.\ref{eq:fincross})
roughly independant of the top mass, when
the $\sqrt{\hat{s}}$ integration range
is larger than the largest technipion width considered. If, however,
  the integration is done over a fixed invariant mass range
smaller than, or comparable to, the width to the lightest
$t\bar{t}$ pair considered,
 the integrated
cross-section decreases with the top mass, since the portion of the width
falling inside the integration bin decreases.

It is essential to estimate the background for this process, in order to see
how clear the signal will be experimentally. The
quantities
 $ \hat{\sigma}_{\,QCD}(gg \rightarrow t\bar{t})$ and
$\hat{\sigma}_{\,QCD}(q\bar{q} \rightarrow t\bar{t})$
\noindent are considered here, as they are expected to give
 the main contribution to the background.
It should be noted  that the technicolor process  has a
substantial interference
with the QCD process $gg \rightarrow t \bar{t}$. This
interference is neglected here, however, since the cuts applied
on the invariant mass (see discussion on cuts below) reduce its effects to
less than $10\%$.
The QCD cross-sections for $q\bar{q}$ and $gg$ fusion
are taken from  Ref.\cite{Combr}, and they are then
inserted into the numerical algorithm, in order to compute
the cross-section $\sigma_{\,bgd}(pp \rightarrow t\bar{t})$.
 For the $q\bar{q}$ processes, the up, down, and
sea-quark distribution functions are used, again according to the
analysis of  Ref.\cite{parton}. The computation
for the Tevatron  takes into account that it is  a $p \bar{p}$ collider,
unlike the SSC and LHC, which will be $pp$ colliders.
Finally, there is an ambiguity as to which scale $\mu$  should be used for the
calculation of the QCD coupling for the background
processes, having s-channel amplitudes on one hand,
and t- and u-channel on the
other, but this does not affect our results by more than $10\%$. The value
$\mu = \sqrt{\hat{s}}$ is used here.

A convenient choice of cuts, $\tau_{\pm}$ for $\tau$, $\pm Y$ for $y_{1}$, and
$y_{\pm}$ for $y_{2}$ is needed in order to reduce  the
background as much as possible.
The minimum possible $\tau \,( = \hat{s}/s)$ cut will be determined by the
experimental resolution
$ \delta \sqrt{ \hat{s}}$ for the invariant mass.
 We will assume that a resolution
of roughly $20$ GeV - approximately $5 \%$
of the $ \sqrt{ \hat{s}}$ values studied
here and comparable to the natural line width
of the technipion - is possible.
This of course depends on the details of detection of the
hadronic jets and charged leptons after both t's undergo the decay
$t \rightarrow Wb $. In Tables 1-3, numerical results are presented for
integrated cross sections for both the technicolor process and the background.
The results are for a $20$-GeV $ \sqrt{ \hat{s}}$ bin centered around the mass
of the technipion. Thus, for this bin,
\begin{equation}
\tau_{\pm}  =  \frac{(M_{P}\;({\rm GeV}) \pm 10\;{\rm GeV} )^{2} }
{s\;({\rm GeV}^{2})}\;.
\end{equation}
\noindent This choice has also the advantage of reducing
 the interference effects of the
technicolor and QCD processes to less than $10\%$, which is satisfactory
for our purposes.

Rapidity cuts are also placed, constraining the longitudinal
momenta of $t$ and $\bar{t}$ to lie between the two wings of a momentum-space
hyperboloid around the beam axis. This reduces
 the QCD background,
 which is larger along the beam axis, and
it also removes from the integration region of Eq.\ref{eq:fincross} part of the
phase space that is experimentally inaccessible, due to its proximity to
the beam axis.
The rapidities $y_{1}$ and $y_{2}$, measured in the laboratory frame, are
constrained by $Y$ and $y_{\pm}$ respectively.
A specific value for $Y$, for a given $\tau$,
automatically determines $y_{\pm}$. The cut $Y = 2.5$ is chosen for
the results presented in Tables 1-3. This
is not very restrictive, since, even for $p_{\perp} = 0$, it allows
$p_{\,\parallel}$ to be roughly as large as $6m_{t}$.
The signal-to-background ratio does not vary substantially with the
choice of Y. This is due to the heaviness of the top quark, which makes the
QCD background less anisotropic than it is for lighter quarks.

\begin{table}[t]
\begin{tabular}{||@{\hspace{-0.9mm}}c@{\hspace{-0.1mm}}
||c|c||c|c||c|c||}\hline
\rule[-3mm]{0cm}{8mm} $M_{P} = 350$ GeV  &
\multicolumn{2}{c||}{$\sqrt{s} = 1.8$ TeV}
& \multicolumn{2}{c||}{$\sqrt{s} = 16$ TeV}&
\multicolumn{2}{c||}{$\sqrt{s} = 40$ TeV}\\  \hline
\rule[-3mm]{0cm}{8mm}$m_{t}$ (GeV)& $\sigma_{\,TC}$ (pb)&
$\sigma_{\,bgd}$ (pb) & $\sigma_{\,TC}$ (pb)&$\sigma_{\,bgd}$ (pb)
&$\sigma_{\,TC}$ (pb)&$\sigma_{\,bgd}$ (pb)\\ \hline \hline
\rule[-3mm]{0cm}{8mm} 90  & 0.24 & 2.8 & 210 & 800 & 870 & 3200
 \\ \hline
\rule[-3mm]{0cm}{8mm} 120 & 0.19 & 2.2 & 160 & 480 & 670 & 1900
 \\ \hline
\rule[-3mm]{0cm}{8mm} 150 & 0.17 & 1.6 & 140 & 210 & 580 & 840
 \\ \hline \hline
\end{tabular}
\caption{ Numerical results for the different cross-sections in picobarns,
 for $M_{P} = 350$ GeV. The numbers given are rounded up to contain
only two significant figures.
The invariant mass bin width is 20 GeV.
The results
are for $N_{TC}=2$;  the technicolor cross-section increases quadratically
with $N_{TC}=2$. Various choices for the top mass are made.
The three different choices for the total C. M.
energy
correspond to the energies of the Tevatron, LHC, and SSC respectively.
The Monte-Carlo relative statistical errors are typically on the
order of $1 \,\%$.}
\end{table}

 From the tables,
it is apparent that colliders with larger center-of-mass energies
lead to larger production cross-sections.
This is because, for a given invariant mass $\sqrt{\hat{s}}$,
they probe regions of smaller $\tau$'s, where the parton,
 and especially the gluon,
distribution functions $f( \sqrt{ \tau}, \hat{s})$ are larger.
Another manifestation of the decrease of $f( \sqrt{ \tau}, \hat{s})$
with increasing
$\tau$ is the fact that, for a given $\sqrt{s}$, larger technipion
masses, corresponding to larger  $\sqrt{\hat{s}}$ values,
give smaller cross-sections.
\begin{table}[t]
\begin{tabular}{||@{\hspace{-0.9mm}}c@{\hspace{-0.1mm}}
||c|c||c|c||c|c||} \hline
\rule[-3mm]{0cm}{8mm} $M_{P} = 450$ GeV  &
\multicolumn{2}{c||}{$ \sqrt{s} = 1.8$ TeV} & \multicolumn{2}{c||}
{$\sqrt{s} = 16$ TeV} &
\multicolumn{2}{c||}
{$\sqrt{s} = 40$ TeV}\\ \hline
\rule[-3mm]{0cm}{8mm}$m_{t}$ (GeV)& $\sigma_{\,TC}$ (pb)&
$\sigma_{\,bgd}$ (pb)
&$\sigma_{\,TC}$ (pb) & $\sigma_{\,bgd}$ (pb)&$\sigma_{\,TC}$ (pb)&
$\sigma_{\,bgd}$ (pb)\\ \hline \hline
\rule[-3mm]{0cm}{8mm}    90  & 0.041 & 0.57 & 100 & 300 & 490 & 1300
 \\ \hline
\rule[-3mm]{0cm}{8mm}    120 & 0.028 & 0.52 & 73  & 210 & 350 & 960
 \\ \hline
\rule[-3mm]{0cm}{8mm}    150 & 0.022 & 0.46 & 54  & 140 & 260 & 640
  \\ \hline
\rule[-3mm]{0cm}{8mm}    180 & 0.019 & 0.39 & 44  & 80  & 210 & 370
\\ \hline \hline
\end{tabular}
\caption{ Results of the same computation as in Table 1, but with
$M_{P} = 450$ GeV. The increased technipion mass allows
larger top masses to be considered. However, results for a top quark heavier
than $180$ GeV are not given, since present constraints on the $ \rho$
 parameter [2] indicate that it is unlikely to be heavier than that.}
\end{table}

For the range of technipion and
top masses chosen, the background gg-fusion process has
a cross-section that decreases with increasing top mass, unlike the $q\bar{q}$
process. Therefore, since the total C.M. energies considered here
 correspond to  small $\tau$ values,
where the gluon distribution functions dominate
over the quark ones, the  total background  decreases
with increasing top mass.
The technicolor cross-sections also
decrease with increasing top-mass, a behavior in
agreement with the rough theoretical expectations
based on Eq.\ref{eq:crossect}.

The Tevatron cross-sections are quite small, and no more than a few signal
events per year should be expected.
The signal-to-background ratio is also
 worse for the Tevatron than  for the
LHC and SSC. This is due to the fact that the $q\bar{q}$ contribution
to the background is larger at the Tevatron, where the $\hat{s}/s$ values
considered  correspond to larger quark distribution functions. Another
reason why the
contribution of $q\bar{q}$ fusion to the background is larger is  that
the Tevatron is a $p\bar{p}$ collider.
At the LHC and SSC,
the technicolor and background cross-sections are comparable to each other,
 indicating that
a  substantial signal could be observed
in future experiments for the range of top masses studied here,
if the $P^{0'}_{8}$ exists.

It is also  useful to
plot the corresponding differential cross-sections in
the invariant mass. The computation of
 $\frac{d \sigma_{\,TC}}{d \sqrt{ \hat{s}}}$
and $\frac{d \sigma_{\,bgd}}{d \sqrt{ \hat{s}}}\;$ can be
 done by a Monte-Carlo algorithm similar to the one used for
 Eq.\ref{eq:fincross}.
 A plot of the differential cross-sections
$\frac{d \sigma_{\,TC}}{d p_{ \perp}}$ and
$\frac{d \sigma_{\,bgd}}{d p_{ \perp}}$
 could be even more useful, as $p_{\perp}$ may be
easier to determine experimentally than $\sqrt{\hat{s}}$ \cite{Ken}. Both of
these plots will appear in a future publication.

\begin{table}[t]
\begin{tabular}{||@{\hspace{-0.9mm}}c@{\hspace{-0.1mm}}
||c@{\hspace{-0.1mm}}|c||c|c||c|c||}\hline
\rule[-3mm]{0cm}{8mm} $M_{P} = 550$ GeV  &
\multicolumn{2}{@{\hspace{-0.1mm}}c||}{$\sqrt{s} = 1.8$ TeV}
& \multicolumn{2}{c||}{$\sqrt{s} = 16$  TeV}&
\multicolumn{2}{c||}{$\sqrt{s} = 40$ TeV}\\ \hline
\rule[-3mm]{0cm}{8mm}$m_{t}$ (GeV)&$\sigma_{\,TC}$ (pb)&
$\sigma_{\,bgd}$ (pb)&$\sigma_{\,TC}$ (pb)&$\sigma_{\,bgd}$ (pb)
&$\sigma_{\,TC}$ (pb)&$\sigma_{\,bgd}$ (pb)\\ \hline \hline
\rule[-3mm]{0cm}{8mm}    90 & 0.78 $ \times 10^{-2}$ & 0.14 &
                               64 & 120 & 310 & 620
 \\ \hline
\rule[-3mm]{0cm}{8mm}   120 & 0.52 $ \times 10^{-2}$ & 0.13 &
                               41 & 96  & 200 & 480
 \\ \hline
\rule[-3mm]{0cm}{8mm}   150 & 0.37 $ \times 10^{-2}$ & 0.12 &
                               29 & 72  & 140 & 360
  \\ \hline
\rule[-3mm]{0cm}{8mm}   180 & 0.29 $ \times 10^{-2}$ & 0.12 &
                               22 & 52  & 110  & 260
 \\ \hline \hline
\end{tabular}
\caption{ Results for $M_{P} = 550$ GeV.}
\end{table}

It is worth pointing out that the signal coming from
 a technipion  $P^{0'}_{8}$
can be  distinguished from a  Higgs particle $H^{0}$,
 since the latter has a much smaller cross-section for the same detection
channel.
Indeed, the fact that the technipion
is a color octet, and that its decay width $\Gamma(P^{0'}_{8} \rightarrow gg)$
depends quadratically on the number of technicolors, makes the technipion
production rate roughly $8 N^{2}_{\,TC}$ times larger than the
corresponding Higgs process \cite{ggtt}.

To conclude, our results show that presently allowed values
for the top mass are such
 that a considerable enhancement of $t \bar{t}$ pairs at LHC and SSC
 energies can be expected from
color-octet technipion production and decay. A similar, but somewhat
smaller, enhancement can be expected from the color-singlet technipion, if it
lies above the $t\bar{t}$ threshold.
 Therefore, the  process considered here can  be used as
a direct test of a large class of technicolor models. Finally, it is
worth noting that the enhancement of $b\bar{b}$ production from
technipion decay could also be interesting, especially if the
technipion masses are below the $t\bar{t}$ production threshold.

\noindent {\bf Acknowledgements} \\
We thank  Charles Baltay, Ken Lane, Steven Manly and Torbjorn Sjostrand
for very helpful
discussions.

\end{document}